\documentclass{article}

\def\noheaderplainsetup{

\topmargin=0pt \headheight=0pt \headsep=0pt  \oddsidemargin=0pt \evensidemargin=0pt  \textheight=8.9truein \textwidth=6.2truein}

\noheaderplainsetup

\usepackage{amsfonts}

\begin{document}


\newcommand{\code}[1]{\ulcorner #1 \urcorner}
\newcommand{\mldi}{\hspace{2pt}\mbox{\footnotesize $\vee$}\hspace{2pt}}
\newcommand{\mlci}{\hspace{2pt}\mbox{\footnotesize $\wedge$}\hspace{2pt}}
\newcommand{\emptyrun}{\langle\rangle} 
\newcommand{\oo}{\bot}            
\newcommand{\pp}{\top}            
\newcommand{\xx}{\wp}               
\newcommand{\legal}[2]{\mbox{\bf Lr}^{#1}_{#2}} 
\newcommand{\win}[2]{\mbox{\bf Wn}^{#1}_{#2}} 
 \newcommand{\one}{\mbox{\sc One}}
 \newcommand{\two}{\mbox{\sc Two}}
 \newcommand{\three}{\mbox{\sc Three}}
 \newcommand{\four}{\mbox{\sc Four}}
 \newcommand{\first}{\mbox{\sc Derivation}}
 \newcommand{\second}{\mbox{\sc Second}}
 \newcommand{\uorigin}{\mbox{\sc Org}}
 \newcommand{\image}{\mbox{\sc Img}}
 \newcommand{\limitset}{\mbox{\sc Lim}}
 \newcommand{\fif}{\mbox{\bf CL15}}
\newcommand{\col}[1]{\mbox{$#1$:}}

\newcommand{\sti}{\mbox{\raisebox{-0.02cm}
{\scriptsize $\circ$}\hspace{-0.121cm}\raisebox{0.08cm}{\tiny $.$}\hspace{-0.079cm}\raisebox{0.10cm}
{\tiny $.$}\hspace{-0.079cm}\raisebox{0.12cm}{\tiny $.$}\hspace{-0.085cm}\raisebox{0.14cm}
{\tiny $.$}\hspace{-0.079cm}\raisebox{0.16cm}{\tiny $.$}\hspace{1pt}}}
\newcommand{\costi}{\mbox{\raisebox{0.08cm}
{\scriptsize $\circ$}\hspace{-0.121cm}\raisebox{-0.01cm}{\tiny $.$}\hspace{-0.079cm}\raisebox{0.01cm}
{\tiny $.$}\hspace{-0.079cm}\raisebox{0.03cm}{\tiny $.$}\hspace{-0.085cm}\raisebox{0.05cm}
{\tiny $.$}\hspace{-0.079cm}\raisebox{0.07cm}{\tiny $.$}\hspace{1pt}}}

\newcommand{\seq}[1]{\langle #1 \rangle}           

\newcommand{\pstb}{\mbox{\raisebox{-0.01cm}{\large $\wedge$}\hspace{-5pt}\raisebox{0.26cm}{\small $\mid$}\hspace{4pt}}}
\newcommand{\pcostb}{\mbox{\raisebox{0.22cm}{\large $\vee$}\hspace{-5pt}\raisebox{0.02cm}{\footnotesize $\mid$}\hspace{4pt}}}

\newcommand{\sst}{\mbox{\raisebox{-0.07cm}{\scriptsize $-$}\hspace{-0.2cm}$\pst$}} 

\newcommand{\scost}{\mbox{\raisebox{0.20cm}{\scriptsize $-$}\hspace{-0.2cm}$\pcost$}} 


\newcommand{\mla}{\mbox{{\Large $\wedge$}}}
\newcommand{\mle}{\mbox{{\Large $\vee$}}}

\newcommand{\pst}{\mbox{\raisebox{-0.01cm}{\scriptsize $\wedge$}\hspace{-4pt}\raisebox{0.16cm}{\tiny $\mid$}\hspace{2pt}}}
\newcommand{\gneg}{\neg}                  
\newcommand{\mli}{\rightarrow}                     
\newcommand{\cla}{\mbox{\large $\forall$}}      
\newcommand{\cle}{\mbox{\large $\exists$}}        
\newcommand{\mld}{\vee}    
\newcommand{\mlc}{\wedge}  
\newcommand{\ade}{\mbox{\Large $\sqcup$}}      
\newcommand{\ada}{\mbox{\Large $\sqcap$}}      
\newcommand{\add}{\sqcup}                      
\newcommand{\adc}{\sqcap}                      

\newcommand{\tlg}{\bot}               
\newcommand{\twg}{\top}               
\newcommand{\st}{\mbox{\raisebox{-0.05cm}{$\circ$}\hspace{-0.13cm}\raisebox{0.16cm}{\tiny $\mid$}\hspace{2pt}}}
\newcommand{\cst}{{\mbox{\raisebox{-0.05cm}{$\circ$}\hspace{-0.13cm}\raisebox{0.16cm}{\tiny $\mid$}\hspace{1pt}}}^{\aleph_0}} 
\newcommand{\cost}{\mbox{\raisebox{0.12cm}{$\circ$}\hspace{-0.13cm}\raisebox{0.02cm}{\tiny $\mid$}\hspace{2pt}}}
\newcommand{\ccost}{{\mbox{\raisebox{0.12cm}{$\circ$}\hspace{-0.13cm}\raisebox{0.02cm}{\tiny $\mid$}\hspace{1pt}}}^{\aleph_0}} 
\newcommand{\pcost}{\mbox{\raisebox{0.12cm}{\scriptsize $\vee$}\hspace{-4pt}\raisebox{0.02cm}{\tiny $\mid$}\hspace{2pt}}}

\newcommand{\psti}{\mbox{\raisebox{-0.02cm}{\tiny $\wedge$}\hspace{-0.121cm}\raisebox{0.08cm}{\tiny $.$}\hspace{-0.079cm}\raisebox{0.10cm}
{\tiny $.$}\hspace{-0.079cm}\raisebox{0.12cm}{\tiny $.$}\hspace{-0.085cm}\raisebox{0.14cm}
{\tiny $.$}\hspace{-0.079cm}\raisebox{0.16cm}{\tiny $.$}\hspace{1pt}}}

\newcommand{\pcosti}{\mbox{\raisebox{0.08cm}{\tiny $\vee$}\hspace{-0.121cm}\raisebox{-0.01cm}{\tiny $.$}\hspace{-0.079cm}\raisebox{0.01cm}
{\tiny $.$}\hspace{-0.079cm}\raisebox{0.03cm}{\tiny $.$}\hspace{-0.085cm}\raisebox{0.05cm}
{\tiny $.$}\hspace{-0.079cm}\raisebox{0.07cm}{\tiny $.$}\hspace{1pt}}}


\newtheorem{theoremm}{Theorem}[section]
\newtheorem{conditionss}{Condition}[section]
\newtheorem{thesiss}[theoremm]{Thesis}
\newtheorem{definitionn}[theoremm]{Definition}
\newtheorem{lemmaa}[theoremm]{Lemma}
\newtheorem{notationn}[theoremm]{Notation}\newtheorem{corollary}[theoremm]{Corollary}
\newtheorem{propositionn}[theoremm]{Proposition}
\newtheorem{conventionn}[theoremm]{Convention}
\newtheorem{examplee}[theoremm]{Example}
\newtheorem{remarkk}[theoremm]{Remark}
\newtheorem{factt}[theoremm]{Fact}
\newtheorem{exercisee}[theoremm]{Exercise}
\newtheorem{questionn}[theoremm]{Open Problem}
\newtheorem{conjecturee}[theoremm]{Conjecture}

\newenvironment{exercise}{\begin{exercisee} \em}{ \end{exercisee}}
\newenvironment{definition}{\begin{definitionn} \em}{ \end{definitionn}}
\newenvironment{theorem}{\begin{theoremm}}{\end{theoremm}}
\newenvironment{lemma}{\begin{lemmaa}}{\end{lemmaa}}
\newenvironment{proposition}{\begin{propositionn} }{\end{propositionn}}
\newenvironment{convention}{\begin{conventionn} \em}{\end{conventionn}}
\newenvironment{remark}{\begin{remarkk} \em}{\end{remarkk}}
\newenvironment{proof}{ {\bf Proof.} }{\  \rule{2.5mm}{2.5mm} \vspace{.2in} }
\newenvironment{idea}{ {\bf Idea.} }{\  \rule{1.5mm}{1.5mm} \vspace{.15in} }
\newenvironment{example}{\begin{examplee} \em}{\end{examplee}}
\newenvironment{fact}{\begin{factt}}{\end{factt}}
\newenvironment{notation}{\begin{notationn} \em}{\end{notationn}}
\newenvironment{conditions}{\begin{conditionss} \em}{\end{conditionss}}
\newenvironment{question}{\begin{questionn}}{\end{questionn}}
\newenvironment{conjecture}{\begin{conjecturee}}{\end{conjecturee}}

\title{The parallel versus branching recurrences in computability logic\thanks{Supported by the NNSF $(60974082)$ of China.}}
\author{Wenyan Xu \ and \ Sanyang Liu}
\date{}
\maketitle

\begin{abstract} This paper shows that the basic logic induced by the parallel recurrence $\pst$ of Computability Logic (i.e., the one in the signature $\{\neg,\wedge,\vee,\pst,\pcost\}$) is a proper superset of the basic logic induced by the branching recurrence $\st$ (i.e., the one in the signature $\{\neg,\wedge,\vee,\st,\cost\}$). The latter is known to be precisely captured by the cirquent calculus system {\bf CL15}, conjectured by Japaridze to remain sound---but not complete---with $\pst$ instead of $\st$. The present result is obtained by positively verifying that conjecture. A secondary result of the paper is showing that $\pst$ is strictly weaker than $\st$ in the sense that, while $\st F$ logically implies $\pst F$, vice versa does not hold.

\end{abstract}


\



\section{Introduction}\label{ssintr}

{\em Computability logic} (CoL), introduced by G. Japaridze \cite{Jap03,Japfin}, is a formal theory of interactive computational problems, understood as games between a machine and its environment (symbolically named as $\top$ and $\bot$, respectively). Formulas in it represent such problems; logical operators stand for operations on them; ``truth" means existence of an algorithmic solution, i.e. $\top$'s effective winning strategy; and validity is understood as truth under every particular interpretation of atoms.

Among the most important operators of CoL are {\em recurrence operators}, in their overall logical spirit reminiscent of the exponentials of linear logic. Recurrences, in turn, come in several flavors, two most natural and basic sorts of which are {\em parallel recurrence} $\pst$ and {\em branching recurrence} $\st$, together with their duals $\pcost,\cost$  defined by $\pcost F=\neg\pst\neg F$ and $\cost F=\neg\st\neg F$. Ample intuitive discussions and elaborations on the two sorts of recurrences and the relations between them were given in \cite{Japtur,Japfour,Japsep,Mosc}. However, finding syntactic characterizations  of the logic induced by recurrences had been among the greatest challenges in CoL until the recent work \cite{Japtam,Japtam2}, where a sound and complete axiomatization, called {\bf CL15},  for the basic $(\neg,\wedge,\vee,\st,\cost)-$fragment of computability logic was constructed.\footnote{The soundness part was proven in \cite{Japtam}, and the completeness part in \cite{Japtam2}.} At the same time, the logical behavior of parallel recurrence $\pst$ still remains largely ununderstood. It is not even known whether the set of principles validated by $\pst$ is recursively enumerable. The present paper brings some initial light into this otherwise completely dark picture. It shows that the set of principles validated by $\pst,\pcost$ in combination with the basic operations $\neg,\wedge,\vee$ is a proper superset of the set of those validated by $\st,\cost$. This is achieved by positively settling Conjecture 6.3 of \cite{Japtam}, according to which {\bf CL15} continues to be sound---but not complete--- with $\pst$ and $\pcost$ instead of $\st$ and $\cost$. Further, to make our investigation of
the relationship between $\pst$ and $\st$ more complete, at the end of the paper we also prove that $\pst$ is strictly weaker than $\st$ in the sense that, while $\st F$ logically implies $\pst F$ (as shown in \cite{Japtur}), vice versa does not hold.



{\bf CL15} is a system built in {\em cirquent calculus}. The latter is a
refinement of sequent calculus. Unlike the more traditional proof theories that manipulate tree-like objects (formulas, sequents, hypersequents, etc.), cirquent calculus deals with graph-style structures called {\em cirquents} (the term is a combination of ``CIRcuit'' and ``seQUENT''), with its main characteristic feature being allowing to explicitly account for possible {\em sharing} of subcomponents between different subcomponents. The approach was introduced by Japaridze
in \cite{Cirq} as a new deductive tool for CoL and was further developed in \cite{Cirdeep,fromto,wenyan1,wenyan2} where a number of advantages of this novel sort of proof theory were revealed, such as high expressiveness, flexibility and efficiency.

In order to make this paper reasonably self-contained, in the next section we reproduce the basic concepts from \cite{Japfin,Japtam} on which the later parts of the paper will rely. An interested reader may consult \cite{Japfin,Japtam} for the associated motivations, detailed explanations and examples.

\section{Preliminaries}

 The letter $\wp$ is used as a variable ranging over $\{\top,\bot\}$, with $\neg\wp$ meaning $\wp$'s adversary. A {\bf move} is a finite string over standard keyboard alphabet. A {\bf labmove} is a move prefixed (``labeled'') with $\top$ or $\bot$. A {\bf run} is a finite or infinite sequence of labmoves, and a {\bf position} is a finite run. Runs are usually delimited by ``$\langle$" and ``$\rangle$", with $\langle\rangle$ thus denoting the {\bf empty run}. For any run $\Gamma$, $\neg\Gamma$ is the same as $\Gamma$, with the only difference that every label $\wp$ is changed to $\neg\wp$.

A {\bf game}\footnote{The concept of a game considered in CoL is more general than the one defined here, with games in the present sense called {\em constant games}. Since we (for simplicity) only consider constant games in this paper, we omit the word ``constant" and just say ``game".} is a pair $A=({\bf Lr}^{A},{\bf Wn}^{A})$, where: (1) ${\bf Lr}^{A}$ is a set of runs satisfying the condition that a finite or infinite run $\Gamma$ is in ${\bf Lr}^{A}$ iff so are all of $\Gamma$'s nonempty finite initial segments.\footnote{This condition can be seen to imply that
the empty run $\langle\rangle$ is always in ${\bf Lr}^{A}$.} If $\Gamma\in{\bf Lr}^{A}$, then $\Gamma$ is said to be a {\bf legal run} of $A$; otherwise $\Gamma$ is an {\bf illegal run} of $A$. A move $\alpha$ is a {\bf legal move} for a player $\wp$ in a position $\Phi$ of $A$ iff $\langle\Phi,\wp\alpha\rangle\in{\bf Lr}^{A}$; otherwise $\alpha$ is an {\bf illegal move}. When the last move of the shortest illegal initial segment of $\Gamma$ is $\wp$-labeled, $\Gamma$ is said to be a $\wp${\bf -illegal run} of $A$. (2) ${\bf Wn}^{A}$ is a function that sends every run $\Gamma$ to one of the players $\top$ or $\bot$, satisfying the condition that if $\Gamma$ is a $\wp$-illegal run of $A$, then ${\bf Wn}^{A}\langle\Gamma\rangle=\neg\wp$. When ${\bf Wn}^{A}\langle\Gamma\rangle=\wp$, $\Gamma$ is said to be a $\wp${\bf -won} run of $A$.\vspace{2mm}

The game operations dealt with in  the present paper are $\neg$ (negation), $\vee$ (parallel disjunction), $\wedge$ (parallel conjunction), $\pst$ (parallel recurrence), $\pcost$ (parallel corecurrence), $\st$ (branching recurrence) and $\cost$ (branching corecurrence). Intuitively, $\neg$ is a role switch operator: $\neg A$ is the game $A$ with the roles of $\top$ and $\bot$ interchanged ($\top$'s legal moves and wins become those of $\bot$, and vice versa). Both $A\wedge B$ and $A\vee B$ are games playing which means playing the two components $A$ and $B$ simultaneously (in parallel). In $A\wedge B$, $\top$ is the winner if it wins in both components, while in $A\vee B$ winning in just one component is sufficient. Next, $\pst A$ is nothing but the infinite parallel conjunction $A\wedge A\wedge A\wedge\ldots$, and $\pcost A$ is nothing but the infinite parallel disjunction $A\vee A\vee A\vee\ldots$. Finally,
both $\st A$ and $\cost A$ are games playing which means simultaneously playing a continuum of copies (or ``threads") of $A$. Each copy/thread is denoted by an infinite bitstring and vice versa, where a {\bf bitstring} is a finite or infinite sequence of bits 0,1.  Making a move $w.\alpha$, where $w$ is a finite bitstring, means making the move $\alpha$ simultaneously in all threads of the form $wy$.  In $\st A$, $\top$ is the winner iff it wins in all threads of $A$, while in $\cost A$ winning in just one thread is sufficient. Again, it should be pointed out that the above is just a very brief and incomplete intuitive characterization. See \cite{Japfin} for more.

Let $\Gamma$ be a run and $\alpha$ be a move. The notation
\begin{center}
$\Gamma^{\alpha}$
\end{center}
will be used to indicate the result of deleting from $\Gamma$ all moves (together with their labels) except those that look like $\alpha\beta$ for some move $\beta$, and then further deleting the prefix ``$\alpha$" from such moves. For instance, $\langle\top 1.\alpha, \bot 2.\beta, \top 1.\gamma, \bot 2.\delta\rangle^{1.}=\langle\top\alpha, \top\gamma\rangle$.

Let $\Omega$ be a run and $x$ be an infinite bitstring. The notation
\begin{center}
$\Omega^{\preceq x}$
\end{center}
will be used to indicate the result of deleting from $\Omega$ all moves (together with their labels) except those that look like $u.\beta$ for some move $\beta$ and some finite initial segment $u$ of $x$, and then further deleting the prefix ``u." from such moves. For instance, $\langle\bot 10.\alpha, \top 111.\beta, \bot 1.\gamma, \bot 00.\alpha\rangle^{\preceq 111\ldots}=\langle\top\beta, \bot\gamma\rangle$.


The earlier-outlined intuitive characterizations of the game operators are captured by the following formal definition.
Below, $A$, $A_1$, $A_2$ are arbitrary games, $\alpha$ ranges over moves, $i\in\{1,2\}$,  $u$ ranges over positive integers identified with its decimal representation, $w$ ranges over finite bitstrings, $x$ ranges over infinite bitstrings, $\Gamma$
is an arbitrary run, and $\Omega$ is any legal run of the game that is being defined.\vspace{2mm}\\
1. $\neg A$ ({\bf negation}) is defined by:\vspace{2mm}

{\bf (i)} $\Gamma\in{\bf Lr}^{\neg A}$ iff $\neg\Gamma\in{\bf Lr}^{A}$.

{\bf (ii)} ${\bf Wn}^{\neg A}\langle\Omega\rangle=\top$ iff ${\bf Wn}^{A}\langle\neg\Omega\rangle=\bot$.\vspace{2mm}\\
2. $A_1\wedge A_2$ ({\bf parallel conjunction}) is defined by:\vspace{2mm}

{\bf (i)} $\Gamma\in{\bf Lr}^{A_1\wedge A_2}$ iff every move of $\Gamma$ is $i.\alpha$ for some $i$,$\alpha$ and, for both $i$, $\Gamma^{i.}\in{\bf Lr}^{A_i}$.

{\bf (ii)} ${\bf Wn}^{A_1\wedge A_2}\langle\Omega\rangle=\top$ iff, for both $i$, ${\bf Wn}^{A_i}\langle\Omega^{i.}\rangle=\top$.\vspace{2mm}\\
3. $A_1\vee A_2$ ({\bf parallel disjunction}) is defined by:\vspace{2mm}

{\bf (i)} $\Gamma\in{\bf Lr}^{A_1\vee A_2}$ iff every move of $\Gamma$ is $i.\alpha$ for some $i$,$\alpha$ and, for both $i$, $\Gamma^{i.}\in{\bf Lr}^{A_i}$.

{\bf (ii)} ${\bf Wn}^{A_1\vee A_2}\langle\Omega\rangle=\top$ iff, for some $i$, ${\bf Wn}^{A_i}\langle\Omega^{i.}\rangle=\top$.\vspace{2mm}\\
4. $\pst A$ ({\bf parallel recurrence}) is defined by:\vspace{2mm}

{\bf (i)} $\Gamma\in \legal{\psti A}{}$ iff every move of $\Gamma$ is $u.\alpha$ for some $u$ and $\alpha$ and, for each such $u$, $\Gamma^{u.}\in\legal{A}{}$.

{\bf (ii)} $\win{\psti A}{}\seq{\Omega}= \pp$ iff, for all $u$, $\win{A}{}\seq{\Omega^{u.}}= \pp$.\vspace{2mm}\\
5. $\pcost A$ ({\bf parallel corecurrence}) is defined by:\vspace{2mm}

{\bf (i)} $\Gamma\in \legal{\pcosti A}{}$ iff every move of $\Gamma$ is $u.\alpha$ for some $u$ and $\alpha$ and, for each such $u$, $\Gamma^{u.}\in\legal{A}{}$.

{\bf (ii)} $\win{\pcosti A}{}\seq{\Omega}= \pp$ iff, for some $u$,  $\win{A}{}\seq{\Omega^{u.}}= \pp$.\vspace{2mm}\\
6. $\st A$ ({\bf branching recurrence})\footnote{The present version of branching (co)recurrence was introduced recently in \cite{Japface}. It is different from yet equivalent to (in all relevant respects) the older version found in \cite{Jap03,Japfin}. The same applies to $\cost$.} is defined by: \vspace{2mm}

{\bf (i)} $\Gamma\in{\bf Lr}^{\sti A}$ iff every move of $\Gamma$ is $w.\alpha$ for some $w$,$\alpha$ and, for all $x$, $\Gamma^{\preceq x}\in{\bf Lr}^{A}$.

{\bf (ii)} ${\bf Wn}^{\sti A}\langle\Omega\rangle=\top$ iff, for all $x$, ${\bf Wn}^{A}\langle\Omega^{\preceq x}\rangle=\top$.\vspace{2mm}\\
7. $\cost A$ ({\bf branching corecurrence}) is defined by: \vspace{2mm}

{\bf (i)} $\Gamma\in{\bf Lr}^{\costi A}$ iff every move of $\Gamma$ is $w.\alpha$ for some $w$,$\alpha$ and, for all $x$, $\Gamma^{\preceq x}\in{\bf Lr}^{A}$.

{\bf (ii)} ${\bf Wn}^{\costi A}\langle\Omega\rangle=\top$ iff, for some $x$, ${\bf Wn}^{A}\langle\Omega^{\preceq x}\rangle=\top$.\vspace{2mm}

In what follows,  we explain---formally or informally---several additional concepts relevant to our proofs.

(1) {\bf Static games}: CoL restricts its attention to a special yet very wide subclass of games termed ``static". Intuitively, static games are interactive tasks where the relative speeds of the players are irrelevant, as it never hurts a player to postpone making moves. A formal definition of this concept can be found in \cite{Japfin}, which we will not reproduce here as nothing in this paper relies on it. The only relevant for us fact, proven in \cite{Jap03,Japfin,Japface}, is that the class of static games is closed under the operations $\neg,\wedge,\vee,\pst,\pcost,\st,\cost$ (as well as any other game operations studied in CoL).

(2) {\bf EPM}: CoL understands $\top$'s effective strategies as interactive machines. Several sorts of such machines have been proposed and studied in CoL, all of them turning out to be equivalent in computing power once we exclusively consider static games. In this paper we only use one  sort of such machines,  called the {\em easy-play machine} ({\bf EPM}). It is a kind of a Turing machine with the additional capability of making moves, and has two tapes\footnote{Often there is also a third tape called the {\em valuation tape}. Its function is to provide values for the variables on which a game may depend. However, as we remember, in this paper we only consider constant games --- games that do not depend on any variables. This makes it possible to safely remove the valuation tape (or leave it there but fully ignore), as this tape is no longer relevant.}: the ordinary read/write {\em work tape}, and the read-only {\em run tape}.  The run tape serves as a dynamic input, at any time (``{\bf clock cycle}") spelling the current position: every time one of the players makes a move, that move---with the corresponding label---is automatically appended to the content of this tape. The machine can make a (one single) move at any time, while its environment can make an (at most one) move only when the machine explicitly allows it to do so (this sort of an action is called {\bf granting permission} ).\footnote{In the more basic sort of machines
 called {\em hard-play machines} ({\bf HPM}), the environment can make any number of moves at any time (needing no ``permission'' for that). It is known (\cite{Jap03,Japfin}) that the two sorts of machines win the same static games.}


(3) {\bf Strategies}:
Let ${\cal M}$ be an EPM. A {\em configuration} of ${\cal M}$ is a full description of the current state of the machine, the contents of its two tapes, and the locations of the corresponding two scanning heads. The {\em initial configuration } is the configuration where ${\cal M}$ is in its start state and both tapes are empty. A configuration $C'$ is said to be an {\em successor} of a configuration $C$ if $C'$ can legally follow $C$ in the standard sense, based on the (deterministic) transition function of the machine and accounting for the possibility of nondeterministic updates of the content of the run tape through environment's moves. A {\bf computation branch} of ${\cal M}$ is a sequence of configurations of ${\cal M}$ where the first configuration is the initial configuration, and each other configuration is a successor of the previous one. Each computation branch $B$ of ${\cal M}$ incrementally spells a run $\Gamma$ on the run tape, which is called the {\bf run spelled by} $B$. Subsequently, any such run $\Gamma$ will be referred to as a {\bf run generated by} ${\cal M}$. A computation branch $B$ of ${\cal M}$ is said to be {\bf fair} iff, in it, permission has been granted infinitely many times. An {\bf algorithmic solution} ({\bf $\top$'s winning strategy}) for a given game $A$ is understood as an EPM ${\cal M}$ such that, whenever $B$ is a computation branch of ${\cal M}$ and $\Gamma$ the run spelled by $B$, $\Gamma$ is a $\top$-won run of $A$, where $B$ should be fair unless $\Gamma$ is a $\bot$-illegal run of $A$. When the above is the case, we say that ${\cal M}$ {\bf wins} $A$.\vspace{2mm}


Now about formulas and the underlying semantics. We have some fixed set of syntactic objects, called {\bf atoms}, for which $P$, $Q$, $R$ will be used as metavariables.
A {\bf formula} is built from atoms in the standard way
using the connectives $\neg$,$\vee$,$\wedge$,$\pst$,$\pcost$,$\st$,$\cost$, with $F\rightarrow G$ understood as an
abbreviation for $\neg F\vee G$ and $\neg$ limited only to atoms, where $\neg\neg F$ is understood as $F$, $\neg(F\wedge G)$ as $\neg F\vee\neg G$, $\neg(F\vee G)$ as $\neg F\wedge\neg G$, $\neg\pst F$ as $\pcost\neg F$, $\neg\pcost F$ as $\pst\neg F$, $\neg\st F$ as $\cost\neg F$, and $\neg\cost F$ as $\st\neg F$. A {\bf $(\neg,\wedge,\vee,\pst,\pcost)$-formula} is one not containing $\st$,$\cost$. Similarly, a {\bf $(\neg,\wedge,\vee,\st,\cost)$-formula} is one not containing $\pst$,$\pcost$. An {\bf interpretation} is a function $^*$ that sends every atom $P$ to a static game $P^*$, and extends to all formulas by seeing the logical connectives as the same-name game operations.
A formula $F$ is {\bf uniformly valid} iff there is an EPM ${\cal M}$, called a {\bf uniform solution} of $F$, such that, for every interpretation $^*$, ${\cal M}$ wins $F^*$.\footnote{Another sort of validity studied in CoL is multiform validity. A formula $F$ is {\bf multiformly valid} iff, for every interpretation $^*$, there is a machine that wins $F^*$. Since uniform validity is stronger than multiform validity, all soundness-style results that we are going to establish about uniform validity automatically extend to multiform validity as well. Partly for this reason, in this paper we will be exclusively interested in uniform validity.}
Throughout the rest of this paper, unless otherwise specified or suggested by the context, by a
``formula'' we will always mean a $(\neg,\wedge,\vee,\pst,\pcost)$-formula.\vspace{2mm}

As noted in section 1, {\bf CL15} is built in {\em cirquent calculus}, whose formalism goes beyond formulas. Namely, a {\bf cirquent} is a triple $C=(\vec{F},\vec{U},\vec{O})$ where: (1) $\vec{F}$ is a nonempty finite sequence of formulas, whose elements are said to be the {\bf oformulas} of $C$. Here the prefix ``o" is used to mean a formula together with a particular occurrence of it in $\vec{F}$. For instance, if $\vec{F}=\langle G, H, H\rangle$, then the cirquent has three oformulas while only two formulas. (2) Both $\vec{U}$ and $\vec{O}$ are nonempty finite sequences of nonempty sets of oformulas of $C$. The elements of $\vec{U}$ are said to be the {\bf undergroups} of $C$, and the elements of $\vec{O}$ are said to be the {\bf overgroups} of $C$. Again, two undergroups (resp. overgroups) may be identical as sets (have identical {\bf contents}), yet they count as different undergroups (resp. overgroups) because they occur at different places in $\vec{U}$ (resp. $\vec{O}$). (3) Additionally, every oformula is required to be in at least one undergroup and at least one overgroup.

Rather than writing cirquents as ordered tuples in the above style, we prefer to represent them through (and identify them with) {\bf diagrams}. Below is such a representation for the cirquent that has four oformulas $E, F, G, H$, three undergroups $\{E,F\}$, $\{F\}$, $\{G,H\}$ and three overgroups $\{E,F,G\}$, $\{G\}$, $\{H\}$.

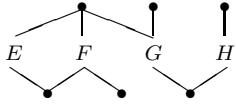
\begin{center}
\begin{picture}(80,50)(0,17)\footnotesize
\put(-10,38){$E\ \ \ \ \ \ \ F\ \ \ \ \ \ \ G\ \ \ \ \ \ \ H$}
\put(-8,35){\line(3,-2){14}}\put(20,35){\line(-3,-2){14}}\put(4,23){$\bullet$}
\put(20,35){\line(3,-2){14}}\put(46,35){\line(3,-2){14}}\put(32,23){$\bullet$}
\put(74,35){\line(-3,-2){14}}\put(58,23){$\bullet$}
\put(19,57){\line(-5,-2){25}}\put(19,57){\line(0,-1){10}}
\put(19,57){\line(5,-2){27}}\put(17,56){$\bullet$}
\put(44,56){$\bullet$}\put(71,56){$\bullet$}
\put(46,57){\line(0,-1){10}}\put(73,57){\line(0,-1){10}}
\end{picture}
\end{center}
Each group in the diagram is represented by (and identified with) a $\bullet$, where the {\bf arcs} (lines connecting the $\bullet$ with oformulas) are pointing to the oformulas that the given group contains.\vspace{2mm}

There are ten inference rules in {\bf CL15}. Below we reproduce those rules from \cite{Japtam} with $\st$ and $\cost$ rewritten as $\pst$ and $\pcost$, respectively. To semantically differentiate the two versions of {\bf CL15} (when necessary), we may use the name ${\bf CL15}(\st)$ for the system that understands (and writes) the recurrence operator as $\st$, and use ${\bf CL15}(\pst)$ for the system that understands (and writes) the recurrence operator as $\pst$.\vspace{2mm}

{\bf Axiom (A):} Axiom is a ``rule" with no premises. It introduces the cirquent

\ \ \ \ \ \ \ $(\langle\neg F_1,F_1,\ldots,\neg F_n,F_n\rangle, \langle\{\neg F_1,F_1\},\ldots,\{\neg F_n,F_n\}\rangle, \langle\{\neg F_1,F_1\},\ldots,\{\neg F_n,F_n\}\rangle)$,\\
where $n$ is any positive integer, and $F_1,\ldots,F_n$ are any formulas. All rules other than Axiom take a single premise.

{\bf Exchange (E):} This rule comes in three versions: {\bf Undergroup Exchange}, {\bf Oformula Exchange} and {\bf Overgroup Exchange}.
The conclusion of Oformula Exchange is obtained by interchanging in the premise two adjacent oformulas $E$ and $F$, and
redirecting to $E$ (resp. $F$) all arcs that were originally pointing to $E$ (resp. $F$). Undergroup (resp. Overgroup) Exchange is the same, with the only difference that the objects interchanged are undergroups (resp. overgroups).

{\bf Duplication (D):} This rule comes in two versions: {\bf Undergroup Duplication} and {\bf Overgroup Duplication}. The conclusion of Undergroup Duplication is obtained by replacing in the premise some undergroup $U$ with two adjacent undergroups whose contents are identical to that of $U$. Similarly for Overgroup Duplication.

{\bf Merging (M):} The conclusion of this rule can be obtained from the premise by merging any two adjacent overgroups $O_1$ and $O_2$ into one overgroup $O$, and including in $O$ all oformulas that were originally contained in $O_1$ or $O_2$ or both.

{\bf Weakening (W):} For the convenience of description, we explain this and the remaining rules in the bottom-up view. The premise of this rule is obtained by deleting in the conclusion an arc between some undergroup $U$ with $\geq 2$ elements and some oformula $F$; if $U$ was the only undergroup containing $F$, then $F$ should also be deleted, together with all arcs between $F$ and overgroups; if such a deletion makes some overgroups empty, then they should also be deleted.

{\bf Contraction (C):} The premise of this rule is obtained by replacing in the conclusion an oformula $\pcost F$ by two adjacent oformulas $\pcost F$ and $\pcost F$, and including both of them in exactly the same undergroups and overgroups in which the original $\pcost F$ was contained.


{\bf Disjunction introduction ($\vee$):} The premise of this rule is obtained by replacing in the conclusion an oformula $E\vee F$ by two adjacent oformulas $E$ and $F$, and including both of them in exactly the same undergroups and overgroups in which the original $E\vee F$ was contained.

{\bf Conjunction introduction ($\wedge$):} According to this rule, if a cirquent (the conclusion) has an oformula $E\wedge F$, then the premise
can be obtained by splitting the original $E\wedge F$ into two adjacent oformulas $E$ and $F$, including both of them in exactly the same overgroups in which the original $E\wedge F$ was contained, and splitting every undergroup $\Gamma$ that originally contained $E\wedge F$ into two adjacent undergroups $\Gamma^{E}$ and $\Gamma^{F}$, where $\Gamma^{E}$ contains $E$ (but not $F$), and $\Gamma^{F}$ contains $F$ (but not $E$), with all other ($\neq E\wedge F$) oformulas of $\Gamma$ contained by both $\Gamma^{E}$ and $\Gamma^{F}$.

{\bf Recurrence introduction ($\pst$):} The premise of this rule is obtained by replacing in the conclusion an oformula $\pst F$ by $F$, with all arcs unchanged, and inserting a new overgroup $\Gamma$ that contains $F$ as its {\em only} oformula.

{\bf Corecurrence introduction ($\pcost$):} The premise of this rule is obtained by replacing in the conclusion an oformula $\pcost F$ by $F$, with all arcs unchanged, and additionally including $F$ in any (possibly zero) number of the already existing overgroups.\vspace{2mm}

Below we provide illustrations for all rules, in each case an abbreviated name of the rule standing next to the horizontal line separating the premise from the conclusion. Our illustration for the axiom (the ``{\bf A}" labeled rule) is a specific cirquent where $n=2$; our illustrations for all other rules are merely examples chosen arbitrarily. Unfortunately, no systematic ways for schematically representing cirquent calculus rules have been elaborated so far. This explains why we appeal to examples instead.

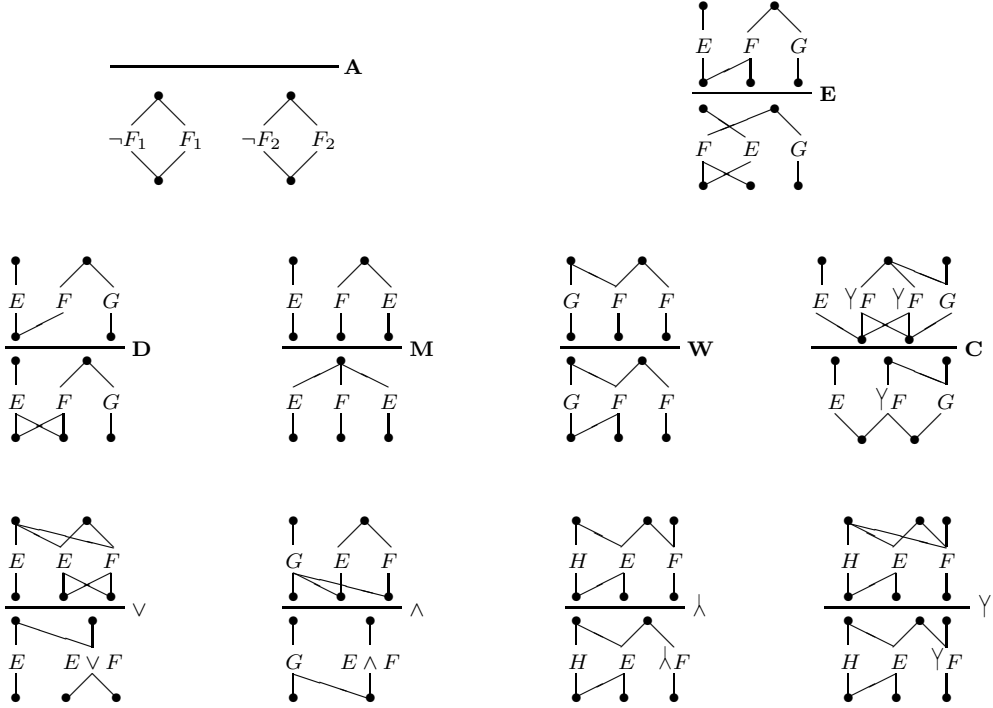
\begin{center}
\begin{picture}(80,90)(60,7)\footnotesize
\put(-70,0){\begin{picture}(80,80)
\put(20,60){\line(1,0){86}}\put(19,30){$\neg F_1\ \ \ \ F_1\ \ \ \ \ \neg F_2\ \ \ \ F_2$}\put(108,57){\bf A}
\put(28,38){\line(1,1){10}}\put(48,38){\line(-1,1){10}}
\put(28,28){\line(1,-1){10}}\put(48,28){\line(-1,-1){10}}
\put(36,15){$\bullet$}\put(36,47){$\bullet$}
\put(78,38){\line(1,1){10}}\put(98,38){\line(-1,1){10}}
\put(78,28){\line(1,-1){10}}\put(98,28){\line(-1,-1){10}}
\put(86,15){$\bullet$}\put(86,47){$\bullet$}
\end{picture}}
\put(40,0){\begin{picture}(80,80)\put(130,50){\line(1,0){45}}\put(178,47){\bf E}
\put(131,65){$E\ \ \ \ F\ \ \ \ G$}\put(134,63){\line(0,-1){8}}\put(152,63){\line(0,-1){8}}\put(170,63){\line(0,-1){8}}
\put(132,52){$\bullet$}\put(150,52){$\bullet$}\put(168,52){$\bullet$}\put(152,63){\line(-2,-1){18}}
\put(132,81){$\bullet$}\put(134,74){\line(0,1){8}}\put(159,81){$\bullet$}\put(161,83){\line(-1,-1){10}}\put(161,83){\line(1,-1){10}}

\put(131,26){$F\ \ \ \ E\ \ \ \ G$}
\put(132,42){$\bullet$}\put(159,42){$\bullet$}
\put(134,44){\line(3,-2){16}}\put(161,44){\line(1,-1){10}}\put(161,44){\line(-5,-2){25}}
\put(132,13){$\bullet$}\put(150,13){$\bullet$}\put(168,13){$\bullet$}
\put(152,24){\line(-2,-1){18}}\put(134,24){\line(0,-1){8}}\put(134,24){\line(2,-1){18}}\put(170,24){\line(0,-1){8}}
\end{picture}}

\end{picture}
\end{center}

\begin{center}
\begin{picture}(80,80)(60,20)\footnotesize

\put(-220,5){\begin{picture}(80,80)\put(130,50){\line(1,0){45}}\put(178,47){\bf D}
\put(0,0){\begin{picture}(80,80)\put(131,65){$E\ \ \ \ F\ \ \ \ G$}\put(134,63){\line(0,-1){8}}\put(170,63){\line(0,-1){8}}
\put(132,52){$\bullet$}\put(168,52){$\bullet$}\put(152,63){\line(-2,-1){18}}
\put(132,81){$\bullet$}\put(134,74){\line(0,1){8}}\put(159,81){$\bullet$}\put(161,83){\line(-1,-1){10}}\put(161,83){\line(1,-1){10}}\end{picture}}

\put(0,-38){\begin{picture}(80,80)\put(131,65){$E\ \ \ \ F\ \ \ \ G$}\put(134,63){\line(0,-1){8}}\put(170,63){\line(0,-1){8}}
\put(132,52){$\bullet$}\put(168,52){$\bullet$}\put(152,63){\line(-2,-1){18}}\put(152,63){\line(0,-1){8}}\put(150,52){$\bullet$}
\put(132,81){$\bullet$}\put(134,74){\line(0,1){8}}\put(159,81){$\bullet$}\put(161,83){\line(-1,-1){10}}\put(161,83){\line(1,-1){10}}
\put(134,63){\line(2,-1){18}}
\end{picture}}
\end{picture}}

\put(-115,5){\begin{picture}(80,80)\put(130,50){\line(1,0){45}}\put(178,47){\bf M}
\put(0,0){\begin{picture}(80,80)\put(131,65){$E\ \ \ \ F\ \ \ \ E$}\put(134,63){\line(0,-1){8}}\put(170,63){\line(0,-1){8}}
\put(132,52){$\bullet$}\put(168,52){$\bullet$}\put(152,63){\line(0,-1){8}}\put(150,52){$\bullet$}
\put(132,81){$\bullet$}\put(134,74){\line(0,1){8}}\put(159,81){$\bullet$}\put(161,83){\line(-1,-1){10}}\put(161,83){\line(1,-1){10}}\end{picture}}

\put(0,-38){\begin{picture}(80,80)\put(131,65){$E\ \ \ \ F\ \ \ \ E$}\put(134,63){\line(0,-1){8}}\put(170,63){\line(0,-1){8}}
\put(132,52){$\bullet$}\put(168,52){$\bullet$}\put(152,63){\line(0,-1){8}}\put(150,52){$\bullet$}
\put(150,81){$\bullet$}\put(152,82){\line(0,-1){8}}\put(152,82){\line(-2,-1){18}}\put(152,82){\line(2,-1){18}}\end{picture}}
\end{picture}}

\put(-10,5){\begin{picture}(80,80)\put(130,50){\line(1,0){45}}\put(178,47){\bf W}
\put(0,0){\begin{picture}(80,80)\put(131,65){$G\ \ \ \ F\ \ \ \ F$}\put(134,63){\line(0,-1){8}}\put(134,82){\line(2,-1){17}}\put(170,63){\line(0,-1){8}}
\put(132,52){$\bullet$}\put(168,52){$\bullet$}\put(152,63){\line(0,-1){8}}\put(150,52){$\bullet$}
\put(132,81){$\bullet$}\put(134,74){\line(0,1){8}}\put(159,81){$\bullet$}\put(161,83){\line(-1,-1){10}}\put(161,83){\line(1,-1){10}}\end{picture}}

\put(0,-38){\begin{picture}(80,80)\put(131,65){$G\ \ \ \ F\ \ \ \ F$}\put(134,63){\line(0,-1){8}}\put(152,63){\line(-2,-1){18}}\put(134,82){\line(2,-1){17}}\put(170,63){\line(0,-1){8}}
\put(132,52){$\bullet$}\put(168,52){$\bullet$}\put(152,63){\line(0,-1){8}}\put(150,52){$\bullet$}
\put(132,81){$\bullet$}\put(134,74){\line(0,1){8}}\put(159,81){$\bullet$}\put(161,83){\line(-1,-1){10}}\put(161,83){\line(1,-1){10}}\end{picture}}
\end{picture}}

\put(90,5){\begin{picture}(80,80)\put(125,50){\line(1,0){55}}\put(183,47){\bf C}
\put(0,0){\begin{picture}(80,80)\put(125,65){$E\ \  \pcost F\ \ \pcost F\ \  G$}\put(127,63){\line(5,-3){18}}\put(144,63){\line(0,-1){9}}\put(162,63){\line(0,-1){9}}\put(178,63){\line(-5,-3){16}}
\put(144,63){\line(2,-1){18}}\put(162,63){\line(-2,-1){18}}
\put(142,51){$\bullet$}\put(160,51){$\bullet$}
\put(127,81){$\bullet$}\put(129,83){\line(0,-1){9}}\put(152,81){$\bullet$}\put(154,83){\line(1,-1){10}}\put(154,83){\line(-1,-1){10}}
\put(154,83){\line(5,-2){22}}
\put(174,81){$\bullet$}\put(176,83){\line(0,-1){9}}\end{picture}}

\put(0,-38){\begin{picture}(80,80)\put(131,65){$E\ \ \ \  \pcost F\ \ \ \ G$}\put(134,63){\line(1,-1){10}}\put(154,63){\line(-1,-1){10}}\put(154,63){\line(1,-1){10}}\put(174,63){\line(-1,-1){10}}
\put(142,51){$\bullet$}\put(162,51){$\bullet$}
\put(132,81){$\bullet$}\put(134,74){\line(0,1){8}}\put(152,81){$\bullet$}\put(154,83){\line(0,-1){9}}\put(154,83){\line(5,-2){22}}
\put(174,81){$\bullet$}\put(176,83){\line(0,-1){9}}
\end{picture}}
\end{picture}}
\end{picture}
\end{center}

\begin{center}
\begin{picture}(80,90)(60,20)\footnotesize

\put(-220,5){\begin{picture}(80,80)\put(130,50){\line(1,0){45}}\put(178,47){\bf $\vee$}
\put(0,0){\begin{picture}(80,80)\put(131,65){$E\ \ \ \ E\ \ \ \ F$}\put(134,63){\line(0,-1){8}}\put(170,63){\line(0,-1){8}}\put(170,63){\line(-2,-1){18}}\put(152,63){\line(2,-1){18}}
\put(132,52){$\bullet$}\put(168,52){$\bullet$}\put(152,63){\line(0,-1){8}}\put(150,52){$\bullet$}
\put(132,81){$\bullet$}\put(134,74){\line(0,1){8}}\put(159,81){$\bullet$}\put(161,83){\line(-1,-1){10}}\put(161,83){\line(1,-1){10}}
\put(133,82){\line(2,-1){18}}\put(133,82){\line(4,-1){38}}\end{picture}}

\put(0,-38){\begin{picture}(80,80)\put(131,65){$E\ \ \ \ \ E\vee F$}\put(134,63){\line(0,-1){8}}\put(163,63){\line(1,-1){10}}

\put(132,52){$\bullet$}\put(170,52){$\bullet$}\put(151,52){$\bullet$}
\put(132,81){$\bullet$}\put(134,74){\line(0,1){8}}\put(161,81){$\bullet$}\put(163,82){\line(0,-1){9}}
\put(163,63){\line(-1,-1){10}}
\put(133,83){\line(3,-1){30}}
\end{picture}}
\end{picture}}

\put(-115,5){\begin{picture}(80,80)\put(130,50){\line(1,0){45}}\put(178,47){\bf $\wedge$}
\put(0,0){\begin{picture}(80,80)\put(131,65){$G\ \ \ \ E\ \ \ \ F$}\put(134,63){\line(0,-1){8}}
\put(134,63){\line(2,-1){18}}\put(134,63){\line(4,-1){36}}\put(170,63){\line(0,-1){8}}
\put(132,52){$\bullet$}\put(168,52){$\bullet$}\put(152,63){\line(0,-1){8}}\put(150,52){$\bullet$}
\put(132,81){$\bullet$}\put(134,74){\line(0,1){8}}\put(159,81){$\bullet$}\put(161,83){\line(-1,-1){10}}\put(161,83){\line(1,-1){10}}\end{picture}}

\put(0,-38){\begin{picture}(80,80)\put(131,65){$G\ \ \ \ \ E\wedge F$}\put(134,63){\line(0,-1){8}}\put(163,63){\line(0,-1){8}}\put(134,63){\line(3,-1){28}}
\put(132,52){$\bullet$}\put(161,52){$\bullet$}
\put(132,81){$\bullet$}\put(134,74){\line(0,1){8}}\put(161,81){$\bullet$}\put(163,83){\line(0,-1){9}}\end{picture}}
\end{picture}}

\put(-8,5){\begin{picture}(80,80)\put(130,50){\line(1,0){45}}\put(178,47){\bf $\pst$}
\put(0,0){\begin{picture}(80,80)\put(131,65){$H\ \ \ \ E\ \ \ \ F$}\put(134,63){\line(0,-1){8}}\put(152,63){\line(-2,-1){18}}\put(134,82){\line(2,-1){17}}\put(171,63){\line(0,-1){8}}
\put(132,52){$\bullet$}\put(169,52){$\bullet$}\put(152,63){\line(0,-1){8}}\put(150,52){$\bullet$}\put(169,81){$\bullet$}\put(171,82){\line(0,-1){9}}
\put(132,81){$\bullet$}\put(134,74){\line(0,1){8}}\put(159,81){$\bullet$}\put(161,83){\line(-1,-1){10}}\put(161,83){\line(1,-1){10}}\end{picture}}

\put(0,-38){\begin{picture}(80,80)\put(131,65){$H\ \ \ \ E\ \ \ \pst F$}\put(134,63){\line(0,-1){8}}\put(152,63){\line(-2,-1){18}}\put(134,82){\line(2,-1){17}}\put(171,63){\line(0,-1){8}}
\put(132,52){$\bullet$}\put(169,52){$\bullet$}\put(152,63){\line(0,-1){8}}\put(150,52){$\bullet$}
\put(132,81){$\bullet$}\put(134,74){\line(0,1){8}}\put(159,81){$\bullet$}\put(161,83){\line(-1,-1){10}}\put(161,83){\line(1,-1){10}}\end{picture}}
\end{picture}}

\put(95,5){\begin{picture}(80,80)\put(125,50){\line(1,0){55}}\put(183,47){\bf $\pcost$}
\put(0,0){\begin{picture}(80,80)\put(131,65){$H\ \ \ \ E\ \ \ \ F$}\put(134,63){\line(0,-1){8}}\put(152,63){\line(-2,-1){18}}\put(134,82){\line(2,-1){17}}\put(171,63){\line(0,-1){8}}\put(134,82){\line(4,-1){37}}
\put(132,52){$\bullet$}\put(169,52){$\bullet$}\put(152,63){\line(0,-1){8}}\put(150,52){$\bullet$}\put(169,81){$\bullet$}\put(171,82){\line(0,-1){9}}
\put(132,81){$\bullet$}\put(134,74){\line(0,1){8}}\put(159,81){$\bullet$}\put(161,83){\line(-1,-1){10}}\put(161,83){\line(1,-1){10}}\end{picture}}

\put(0,-38){\begin{picture}(80,80)\put(131,65){$H\ \ \ \ E\ \ \ \pcost F$}\put(134,63){\line(0,-1){8}}\put(152,63){\line(-2,-1){18}}\put(134,82){\line(2,-1){17}}
\put(171,63){\line(0,-1){8}}\put(169,81){$\bullet$}\put(171,82){\line(0,-1){9}}
\put(132,52){$\bullet$}\put(169,52){$\bullet$}\put(152,63){\line(0,-1){8}}\put(150,52){$\bullet$}
\put(132,81){$\bullet$}\put(134,74){\line(0,1){8}}\put(159,81){$\bullet$}\put(161,83){\line(-1,-1){10}}\put(161,83){\line(1,-1){10}}\end{picture}}
\end{picture}}
\end{picture}
\end{center}

\vspace{2mm}
The above are all ten rules of {\bf CL15$(\pst)$}. A {\bf CL15$(\pst)$-proof} (or simply a {\bf proof}) of a cirquent $C$ is a sequence $\langle C_1,\ldots,C_n\rangle$ of cirquents, where $n\geq 1$, such that $C_n=C$, $C_1$ is an axiom, and $C_{i}$ ($1< i\leq n$) follows from $C_{i-1}$ by one of the rules of {\bf CL15$(\pst)$}. For any formula $F$, the expression $F^{\clubsuit}$ is used to denote the cirquent $(\langle F\rangle,\langle\{F\}\rangle,\langle\{F\}\rangle)$. Then a {\bf CL15$(\pst)$-proof} (or simply a {\bf proof}) of a formula $F$ is stipulated to be  a proof of the cirquent $F^{\clubsuit}$. A formula or cirquent $X$ is {\bf provable}, symbolically {\bf CL15}$(\pst)\vdash X$, iff it has a proof.\vspace{2mm}

As mentioned,  {\bf CL15}$(\st)$ is the same as {\bf CL15}$(\pst)$, only with $\st$,$\cost$ instead of $\pst$,$\pcost$.

\begin{theorem}\label{niu}
{\em (Japaridze \cite{Japtam,Japtam2})} A $(\gneg,\mlc,\mld,\st,\cost)$-formula is uniformly valid iff it is provable in {\bf CL15}$(\st)$.
\end{theorem}

\section{A semantics of cirquents}
To prove the soundness of {\bf CL15}$(\pst)$, we need to extend the earlier-described semantics from formulas to cirquents.

\begin{notation}
Let $\Gamma$ be a run, $a$ be a positive integer, and $\vec{x}=x_1,\ldots,x_n$ be a nonempty sequence of $n$ {\em positive integers}. We will be using the notation
\begin{center}
$\Gamma^{[a;\vec{x}]}$
\end{center}
to mean the result of
\begin{itemize}
\item deleting from $\Gamma$ all moves (together with their labels) except those that look like $a;u_1,\ldots,u_n.\beta$ for some move $\beta$ and some sequence of $n$ {\em natural numbers} $u_1,\ldots,u_n$ satisfying the condition that whenever $u_i\neq 0$ $(i\in\{1,\ldots,n\})$, $u_i=x_i$, and
\item then further deleting the prefix ``$a;u_1,\ldots,u_n.$" from such moves.
\end{itemize}
For instance, $\langle\bot 1;1,1.\alpha, \top 1;1,2.\beta, \bot 1;1,0.\gamma, \bot 2;1,0.\delta\rangle^{[1;1,2]}=\langle\top\beta, \bot\gamma\rangle$.
\end{notation}

\begin{definition}
Let $^*$ be an interpretation, and $C=(\langle F_1,\ldots,F_k\rangle,\langle U_1,\ldots,U_m\rangle,\langle O_1,\ldots,O_n\rangle)$ be a cirquent. Then $C^*$ is the game defined as follows, where $\Gamma$ is an arbitrary run and $\Omega$ is any legal run of $C^*$.\vspace{2mm}\\
{\bf (i)} $\Gamma\in {\bf Lr}^{C^*}$ iff the following two conditions are satisfied:
\begin{itemize}
\item Every move of $\Gamma$ looks like $a;\vec{u}.\alpha$, where $\alpha$ is some move, $a\in\{1,\ldots,k\}$, and $\vec{u}=u_1,\ldots,u_n$ is a sequence of $n$ natural numbers such that, for every $j\in\{1,\ldots,n\}$, we have $u_j=0$ iff the overgroup $O_j$ does not contain the oformula $F_a$.
\item For every $a\in\{1,\ldots,k\}$ and every sequence $\vec{x}$ of $n$ positive integers, $\Gamma^{[a;\vec{x}]}\in {\bf Lr}^{F_a^{*}}$.
\end{itemize}
{\bf (ii)} ${\bf Wn}^{C^*}\langle\Omega\rangle=\top$ iff, for every $i\in\{1,\ldots,m\}$ and every sequence $\vec{x}$ of $n$ positive integers, there is an $a\in\{1,\ldots,k\}$ such that the undergroup $U_i$ contains the oformula $F_a$ and ${\bf Wn}^{F_a^*}\langle\Omega^{[a;\vec{x}]}\rangle=\top$.
\end{definition}

\begin{remark}\label{feb13a}
Intuitively, any legal run $\Omega$ of $C^*$ consists of parallel plays of countably infinite copies of each of the games $F_{a}^{*}$ ($1\leq a\leq k$). To every sequence $\vec{x}$ of $n$ positive integers corresponds a copy of $F_a^*$, and $\Omega^{[a;\vec{x}]}$ is the run played in that copy. We shall simply say {\bf the copy $\vec{x}$} of $F_a^*$ to mean the copy of $F_a^*$ which corresponds to the sequence $\vec{x}$. Now, consider a given undergroup $U_i$. $\top$ is the winner in $U_i$ iff, for every sequence $\vec{x}$ of $n$ positive integers, there is an oformula $F_a$ in $U_i$ such that $\Omega^{[a;\vec{x}]}$ is won by $\pp$. Finally, $\pp$ wins the overall game $C^*$ iff it wins in all undergroups of $C$. In fact, overgroups can be seen as generalized $\pst$s, with the only main difference that the former can be shared by several oformulas; undergroups can be seen as generalized disjunctions, with the only main difference that the former may have shared arguments with other undergroups.

\end{remark}

We say that a cirquent $C$ is
{\bf uniformly valid} iff there is an EPM $\cal M$, called a {\bf uniform solution} of $C$, such that, for every interpretation $^*$, $\cal M$ wins $C^*$.

\section{Main results}

\begin{lemma}\label{apr14a}
There is an effective function $f$ from EPMs to EPMs such that, for every EPM ${\cal M}$, formula $F$ and interpretation $^*$, if ${\cal M}$ wins $\pst F^*$, then $f({\cal M})$ wins $F^*$.
\end{lemma}

\begin{proof}
Our proof here almost literally follows the proof of Lemma 9.1 of \cite{Japtam}. It is known that affine logic proves $\pst P\rightarrow P$. At the same time, according to Theorem 37 of \cite{Japfin}, affine logic is sound with respect to uniform validity. So, the formula $\pst P\rightarrow P$ is uniformly valid. This almost immediately implies that there is an EPM ${\cal N}_0$ such that ${\cal N}_0$ wins $\pst F^*\rightarrow F^*$ for any formula $F$ and interpretation $^*$. Furthermore, by Proposition 21.3 of \cite{Jap03}, there is an effective procedure that, for any pair $({\cal N},{\cal M})$ of EPMs, returns an EPM $h({\cal N},{\cal M})$ such that, for any static games $A$ and $B$, if ${\cal N}$ wins $A\rightarrow B$ and ${\cal M}$ wins $A$, then $h({\cal N},{\cal M})$ wins $B$. So, let $f({\cal M})$ be the function satisfying  $f({\cal M})=h({\cal N}_0,{\cal M})$. Then $f({\cal M})$ wins $F^*$.
\end{proof}

\begin{lemma}\label{apr14b}
There is an effective function $g$ from EPMs to EPMs such that, for every EPM ${\cal M}$, formula $F$ and interpretation $^*$, if ${\cal M}$ wins $(F^{\clubsuit})^*$, then $g({\cal M})$ wins $F^*$.
\end{lemma}

\begin{proof}
Again, it should be acknowledged that the present proof very closely follows
the proof of Lemma 9.2 of \cite{Japtam}, even though there are certain differences.

Every legal move of $(F^{\clubsuit})^*$ looks like $1;u.\alpha$ for some positive integer $u$ and move $\alpha$, while the corresponding legal move of $(\pst F)^*$ simply looks like $u.\alpha$, and vice versa. Consider an arbitrary EPM ${\cal M}$ and an arbitrary  interpretation $^*$. Below we  show the existence of an effective function $f$ such that, if ${\cal M}$ wins $(F^{\clubsuit})^*$, then (the strategy) $f({\cal M})$ wins $(\pst F)^*$.

We construct an EPM $f({\cal M})$ that plays $(\pst F)^*$ by simulating and mimicking a play of $(F^{\clubsuit})^*$ (called the {\bf imaginary play}) by ${\cal M}$ as follows. Throughout simulation, $f({\cal M})$ grants permission whenever the simulated ${\cal M}$ does so, and feeds its environment's response---in a slightly modified form described below---back to the simulated $\cal M$ as the response of ${\cal M}$'s imaginary adversary (this detail of simulation will no longer be explicitly mentioned later in similar situations).  Whenever the environment makes a move $u.\alpha$ for some positive integer $u$ and move $\alpha$, $f({\cal M})$  translates it as the move $1;u.\alpha$ made by the imaginary adversary of ${\cal M}$, and ``vice versa":
whenever the simulated ${\cal M}$ makes a move $1;u.\alpha$ for some positive integer $u$ and move $\alpha$ in the imaginary play of $(F^{\clubsuit})^*$, $f({\cal M})$ translates it as its own move $u.\alpha$ in the real play of $(\pst F)^*$.
The effect achieved by $f({\cal M})$'s strategy can be summarized by saying that it synchronizes every copy of $F^*$ in the real play of $(\pst F)^*$ with the ``same copy" of $F^*$ in the imaginary play of $(F^{\clubsuit})^*$.

Let $\Gamma$ be an arbitrary run generated by $f({\cal M})$, and $\Omega$ be the corresponding run in the imaginary play of $(F^{\clubsuit})^*$ by ${\cal M}$. From our description of  $f({\cal M})$ it is clear that the latter never makes illegal moves unless its environment or the simulated ${\cal M}$ does so first. Hence we may safely assume that $\Gamma$ is a legal run of $(\pst F)^*$ and $\Omega$ is a legal run of $(F^{\clubsuit})^*$, for otherwise either $\Gamma$ is a $\bot$-illegal run of $(\pst F)^*$ and thus $f({\cal M})$ is an automatic winner in $(\pst F)^*$, or $\Omega$ is a $\top$-illegal run of $(F^{\clubsuit})^*$ and thus ${\cal M}$ does not win $(F^{\clubsuit})^*$. Now, it is not hard to see that, for any positive integer $x$, we have $\Gamma^{x.}=\Omega^{[1;x]}$. Therefore, $f({\cal M})$ wins $(\pst F)^*$ as long as ${\cal M}$ wins $(F^{\clubsuit})^*$.

Finally, in view of Lemma \ref{apr14a}, the existence of function $g$ satisfying the promise of the present lemma is obviously guaranteed.
\end{proof}

A rule of {\bf CL15}$(\pst)$ (other than Axiom) is said to be {\bf uniform-constructively sound} iff there is an effective procedure that takes any instance $(A,B)$ (i.e. a particular premise-conclusion pair) of the rule, any EPM ${\cal M}_A$ and returns an EPM ${\cal M}_B$ such that, for any interpretation $^*$, whenever ${\cal M}_A$ wins $A^*$, ${\cal M}_B$ wins $B^*$. Axiom is uniform-constructively sound iff there is an effective procedure that takes any instance $B$ of (the ``conclusion" of) Axiom and returns a uniform solution ${\cal M}_B$ of $B$.

\begin{theorem}\label{mainth1}
All rules  of {\bf CL15}$(\pst)$ are uniform-constructively sound.
\end{theorem}

\begin{proof}
 In what follows, $A$ is the premise of an arbitrary instance of a given rule of {\bf CL15}$(\pst)$, and $B$ is the corresponding conclusion, except the case of Axiom where we only have $B$. We will prove that each rule of {\bf CL15}$(\pst)$ is uniform-constructively sound by showing that  an EPM  ${\cal M}_B$ can be constructed effectively from an arbitrary EPM ${\cal M}_A$ such that, for whatever interpretation $^*$,
  whenever ${\cal M}_A$ wins $A^{\ast}$,  ${\cal M}_B$ wins $B^{\ast}$. Since an interpretation $^{\ast}$ is never relevant in such proofs, we may safely omit it, writing simply $A$ instead of $A^{\ast}$ to represent a game. Next, in all cases the assumption that ${\cal M}_A$ wins $A$ will be implicitly made, even though it should be pointed out that the construction of ${\cal M}_B$ never depends on this assumption. Correspondingly, it will be assumed that ${\cal M}_A$ never makes illegal moves. Further, as in the proof of Lemma \ref{apr14b}, we shall always implicitly assume that ${\cal M}_B$'s adversary never makes illegal moves either. To summarize, when analyzing ${\cal M}_B$, ${\cal M}_A$ and the games they play, we safely pretend that illegal runs never occur.\vspace{2mm}

{\bf (1)}\ Assume that $B$ is an axiom with $2n$ oformulas. An EPM ${\cal M}_B$ that wins $B$ can be constructed as follows. It keeps granting permission. Whenever the environment makes a move $a;\vec{w}.\alpha$, where $1\leq a\leq 2n$ and $\vec{w}$ is a sequence of $n$ natural numbers, ${\cal M}_B$ responds by the move $b;\vec{w}.\alpha$, where $b=a+1$ if $a$ is odd, and $b=a-1$ if $a$ is even. Then, for any run $\Gamma_B$ of $B$ generated by ${\cal M}_B$ and any sequence $\vec{x}$ of $n$ positive integers , we have $\Gamma_B^{[a;\vec{x}]}=\neg\Gamma_B^{[b;\vec{x}]}$. It is obvious that $\Gamma_B$ is a $\top$-won run of $B$, so that ${\cal M}_B$ wins $B$. \vspace{2mm}

{\bf (2)}\ Assume that $B$ follows from $A$ by Overgroup Exchange, where the $i$'th ($i\geq 1$) and the $(i+1)$'th overgroups of $A$ have been swapped when obtaining $B$ from $A$. The EPM ${\cal M}_B$ works by simulating and mimicking ${\cal M}_A$ as follows. Let $n$ be the number of overgroups of either cirquent, and $a$ be a positive integer not exceeding the number of oformulas of either cirquent. For any move (by either player) $a;\vec{w_1},u_1,u_2,\vec{w_2}.\alpha$ of the real play of $B$,  where $\vec{w_1}$ and $\vec{w_2}$ are any sequences of $i-1$ and $n-i-1$ natural numbers, respectively, and $u_1,u_2$ are two natural numbers,  ${\cal M}_B$ translates it as the move $a;\vec{w_1},u_2,u_1,\vec{w_2}.\alpha$ (by the same player) of the imaginary play of $A$, and vice versa, with all other moves not reinterpreted.  Let $\Gamma_B$ be any run generated by ${\cal M}_B$, and $\Gamma_A$ be the corresponding imaginary run generated by ${\cal M}_A$. It is obvious that, for any sequence $\vec{x}$ of $n$ positive integers, $\Gamma_B^{[a;\vec{x}]}=\Gamma_A^{[a;\vec{y}]}$, where $\vec{y}$ is the result of swapping in $\vec{x}$ the $i$'th and $(i+1)$'th integers. Hence ${\cal M}_B$ wins $B$ (because ${\cal M}_A$ wins $A$).

In the case of Oformula Exchange, a similar method can be used to construct ${\cal M}_B$, with the only difference that the reinterpreted objects are the occurrences of two adjacent oformulas rather than the occurrences of two adjacent overgroups.

As for Undergroup Exchange, its conclusion, as a game, is the same as its premise. So, the machine ${\cal M}_B={\cal M}_A$ does the job.\vspace{2mm}

In the subsequent clauses, as in the preceding one, without any further indication, $\Gamma_B$ will stand for an arbitrary run of $B$ generated by ${\cal M}_B$, and $\Gamma_A$ will stand for the run of $A$ generated by the simulated machine ${\cal M}_A$ in the corresponding scenario.\vspace{2mm}

{\bf (3)}\ Assume $B$ is obtained from $A$ by Weakening. If no oformula of $B$ was deleted when moving from $B$ to $A$, then ${\cal M}_B$ works exactly as ${\cal M}_A$ does and succeeds, because every $\top$-won run of $A$ is also a $\top$-won run of $B$ (but not necessarily vice versa). If, when moving from $B$ to $A$, an oformula $F_a$ of $B$ was deleted, then ${\cal M}_B$ can be constructed as a machine that works by simulating and mimicking ${\cal M}_A$. What ${\cal M}_B$ needs to do during its work is to ignore the moves within $F_a$, and play exactly as ${\cal M}_A$ does in all other oformulas. Again, it is obvious that every $\top$-won run of $A$ is also a $\top$-won run of $B$, which means that ${\cal M}_B$ wins $B$ as long as ${\cal M}_A$ wins $A$.\vspace{2mm}

{\bf (4)}\ Since Exchange has already been proven to be uniform-constructively sound, in this and the remaining clauses of the present proof, we may safely assume that the oformulas and overgroups affected by a rule are at the end of the corresponding lists of objects of the corresponding cirquents.

Assume $B$ follows from $A$ by Contraction, and the contracted oformula $\pcost F$ is at the end of the list of oformulas of $B$. Let $a$ be the number of oformulas of $B$, and let $b=a+1$. Thus, the $a$'th oformula of $B$ is $\pcost F$, and the $a$'th and $b$'th oformulas of $A$ are $\pcost F$ and $\pcost F$. Next, let $n$ be the number of overgroups in either cirquent.
As always, we let ${\cal M}_B$ be an EPM that works by simulating and mimicking ${\cal M}_A$. Namely, let $\vec{w}$ be any sequence of $n$ natural numbers. If the moves take place within the oformulas other than $\pcost F$, then nothing should be reinterpreted. If the moves take place in $\pcost F$, then we have:

\begin{itemize}
  \item For any move $a;\vec{w}.u.\alpha$ (by either player) in the real play of $B$, where $u=2k-1$ for some $k\in\{1,2,3,\ldots\}$, ${\cal M}_B$ translates it as the move $a;\vec{w}.k.\alpha$ (by the same  player) of the imaginary play of $A$, and vice versa.
  \item For any move $a;\vec{w}.v.\alpha$ (by either player) in the real  play of $B$, where $v=2m$ for some $m\in\{1,2,3,\ldots\}$, ${\cal M}_B$ translates it as the move $b;\vec{w}.m.\alpha$ (by the same player) of the imaginary play of $A$, and vice versa.
\end{itemize}

Below we will show that ${\cal M}_B$ wins $B$, i.e., ${\cal M}_B$ is the winner in every undergroup of $B$.
Let $U_i^{B}$ be any $i$'th undergroup of $B$ and $U_i^{A}$ be the corresponding $i$'th undergroup of $A$, and let $\vec{x}$ be any sequence of $n$ positive integers. Since ${\cal M_A}$ wins $A$, $U_i^{A}$ is won by ${\cal M}_A$. So, for the sequence $\vec{x}$, there is an oformula $F_j$ ($1\leq j\leq b$) in $U_i^{A}$ such that $\Gamma_A^{[j;\vec{x}]}$ is a $\top$-won run of $F_j$. Next, if such $F_j$ is not one of the two contracted oformulas $\pcost F$ and $\pcost F$, then, for $\vec{x}$, the corresponding oformula $F_j$ of $B$ is also won by ${\cal M}_B$, i.e. $\Gamma_B^{[j;\vec{x}]}$ is a $\top$-won run of $F_j$, because ${\cal M}_B$ plays in the copy $\vec{x}$ of $F_j$ exactly as ${\cal M}_A$ does. This means that $U_i^{B}$ is won by ${\cal M}_B$. If such $F_j$ is one of the two contracted oformulas $\pcost F$ and $\pcost F$, below let us assume that $F_j$ is the left $\pcost F$, with the case of the right $\pcost F$ being similar. Then there is a positive integer $w$ such that the $w$'th component $F$ of the copy $\vec{x}$ of the left $\pcost F$ is won by ${\cal M}_A$, i.e. $(\Gamma_A^{[j;\vec{x}]})^{w.}$ is a $\top$-won run of $F$. But, according to the above description, ${\cal M}_B$ plays in the $(2w-1)$'th component $F$ of the copy $\vec{x}$ of $\pcost F$ in $B$ exactly as ${\cal M}_A$ plays in the $w$'th component $F$ of the copy $\vec{x}$ of the left $\pcost F$ in $A$, i.e. $(\Gamma_B^{[j;\vec{x}]})^{(2w-1).}=(\Gamma_A^{[j;\vec{x}]})^{w.}$. Therefore, $(\Gamma_B^{[j;\vec{x}]})^{(2w-1).}$ is a $\top$-won run of $F$, which means that $\Gamma_B^{[j;\vec{x}]}$ is a $\top$-won run of $\pcost F$ in $B$, and hence the $\pcost F$-containing undergroup $U_i^{B}$ is won by ${\cal M}_B$.\vspace{2mm}

{\em Remark}\hspace{1pt}:\  In the remaining clauses, just as in the preceding one, when talking about playing, winning, etc. in $A$ (resp. $B$) or any of its components, it is to be understood in the context of $\Gamma_A$ (resp. $\Gamma_B$). Furthermore, if $A$ and $B$ have the same number $n$ of overgroups, then the context will additionally include some arbitrary but fixed sequence $\vec{x}$ of $n$ positive integers.\vspace{2mm}

{\bf (5)}\ Undergroup Duplication does not modify the game associated with the cirquent, so we only need to consider Overgroup Duplication.

Assume $B$ is obtained from $A$ by Overgroup Duplication. We assume that the duplicated overgroup is at the end of the list of overgroups of $A$. Let $n+1$ be the number of overgroups of $A$. Thus, every legal move of $A$ (resp. $B$) looks like $a;\vec{w},u.\alpha$ (resp. $a;\vec{w},u_1,u_2.\alpha$), where $a$ is a positive integer not exceeding the number of oformulas of $A$, $\vec{w}$ is a sequence of $n$ natural numbers, and $u,u_1,u_2$ are natural numbers.

Let $f$ be some standard 1-to-1 correspondence from the set of all pairs of positive integers to the set of all positive integers.
As before, ${\cal M}_B$ works by simulating ${\cal M}_A$. Whenever ${\cal M}_A$ makes a move $a;\vec{w},0.\alpha$ in $A$,  ${\cal M}_B$ makes the move $a;\vec{w},0,0.\alpha$ in the real play of $B$, and vice versa. Whenever ${\cal M}_A$ makes the move $a;\vec{w},u.\alpha$ in $A$ for some positive integer $u$, ${\cal M}_B$ makes the move $a;\vec{w},u_1,u_2.\alpha$ in $B$, where $u_1,u_2$ are integers with $f(u_1,u_2)=u$, and vice versa. Note that ${\cal M}_A$'s (legally) making a move $a;\vec{w},0.\alpha$ means that the $a$'th oformula $F_a$ of $A$ is not contained in the $(n+1)$'th overgroup $O_{n+1}$ that was duplicated when moving from $A$ to $B$, which, in turn, means that the corresponding $F_a$ of $B$ is contained in neither the $(n+1)$'th overgroup $O'_{n+1}$ nor the $(n+2)$'th overgroup $O'_{n+2}$ of $B$. Similarly, if ${\cal M}_A$ makes a move $a;\vec{w},u.\alpha$ for some positive integer $u$, then $F_a$ is contained in $O_{n+1}$ of $A$, and hence the corresponding $F_a$ of $B$ is contained in both $O'_{n+1}$ and $O'_{n+2}$ of $B$, with the case of $F_a$ being contained in $O'_{n+1}$ but not in $O'_{n+2}$ (or in $O'_{n+2}$ but not in $O'_{n+1}$) being impossible.

For every oformula $F_a$ of either cirquent, every sequence $\vec y$ of $n$ positive integers and any positive integers $x_1$ and $x_2$, we have $\Gamma_B^{[a;\vec{y},x_1,x_2]}=\Gamma_A^{[a;\vec{y},x]}$, where $x=f(x_1,x_2)$. So it is obvious that ${\cal M}_B$ wins $B$ as long as ${\cal M}_A$ wins $A$.\vspace{2mm}

{\bf (6)}\ Assume $B$ follows from $A$ by Merging. Let us assume that $A$ has $n+2$ overgroups, and $B$ is the result of merging in $A$ the two adjacent overgroups $O_{n+1}$ and $O_{n+2}$. Then every legal move of $A$ (resp. $B$) looks like $a;\vec{w},u_1,u_2.\alpha$ (resp. $a;\vec{w},u.\alpha$), where $a$ is a positive integer not exceeding the number of oformulas in either cirquent, $\vec{w}$ is a sequence of $n$ natural numbers, and $u,u_1,u_2$ are natural numbers. 
The EPM ${\cal M}_B$ works as follows.

If the $a$'th oformula of $A$ is neither in $O_{n+1}$ nor in $O_{n+2}$, then ${\cal M}_B$ interprets every move $a;\vec{w},0,0.\alpha$ made by ${\cal M}_A$ in the imaginary play of $A$ as the move $a;\vec{w},0.\alpha$ in the real play of $B$, and vice versa.

If the $a$'th oformula of $A$ is in $O_{n+1}$ but not in $O_{n+2}$, ${\cal M}_B$ interprets every move $a;\vec{w},v,0.\alpha$ ($v$ is a positive integer) made by ${\cal M}_A$ in the imaginary play of $A$ as the move $a;\vec{w},v.\alpha$ that ${\cal M}_B$ itself should make in the real play of $B$, and vice versa. Namely, ${\cal M}_B$ interprets every move $a;\vec{w},v.\alpha$ by its environment in the real play of $B$ as the move $a;\vec{w},v,0.\alpha$ by ${\cal M}_A$'s adversary in the imaginary play of $A$.

The case of the $a$'th oformula of $A$ being in $O_{n+2}$ but not in $O_{n+1}$ is similar.

Now assume that the $a$'th oformula of $A$ is in both $O_{n+1}$ and $O_{n+2}$. ${\cal M}_B$ interprets every move $a;\vec{w},v_1,v_2.\alpha$ by ${\cal M}_A$ in the imaginary play of $A$ as the move $a;\vec{w},v.\alpha$ in the real play of $B$, where $v_1,v_2,v$ are positive integers such  that $v=f(v_1,v_2)$, with $f$ here standing for the pairing function explained in the preceding clause of this proof.

For every oformula $F_a$ of either cirquent, every sequence $\vec y$ of $n$ positive integers and any positive integer $x$, we have $\Gamma_B^{[a;\vec{y},x]}=\Gamma_A^{[a;\vec{y},x_1,x_2]}$, where $x_1,x_2$ are positive integers satisfying that $x_1=x$ (when $F_a$ is contained in $O_{n+1}$ but not $O_{n+2}$), or $x_2=x$ (when $F_a$ is contained in $O_{n+2}$ but not $O_{n+1}$), or $f(x_1,x_2)=x$ (when $F_a$ is contained in both $O_{n+1}$ and $O_{n+2}$, or is contained in neither of them). So it is obvious that ${\cal M}_B$ wins $B$ as long as ${\cal M}_A$ wins $A$.\vspace{2mm}

{\bf (7)}\ In this and the remaining clauses of the present proof, we will limit our descriptions to what moves ${\cal M}_B$ needs to properly reinterpret and how, with any unmentioned sorts of moves implicitly assumed to remain unchanged.

Assume $B$ is obtained from $A$ by Disjunction Introduction. Let us assume that the last ($a$'th) oformula of $B$ is $E\vee F$, and the last two ($a$'th and $b$'th, where $b=a+1$) oformulas of $A$ are $E$ and $F$. We let ${\cal M}_B$ reinterpret every move $a;\vec{w}.\alpha$ (resp. $b;\vec{w}.\alpha$) by either player in the imaginary play of $A$ as the move $a;\vec{w}.1.\alpha$ (resp. $a;\vec{w}.2.\alpha$) by the same player in the real play of $B$, and vice versa.

Consider any undergroup $U_i^{B}$ of $B$, and let $U_i^{A}$ be the corresponding undergroup of $A$.  As before, ${\cal M}_A$'s winning $A$ means that $U_i^{A}$ is won by ${\cal M}_A$, which, in turn, means that there is an oformula $G$ in $U_i^{A}$ that is won by ${\cal M}_A$. If $G$ is neither $E$ nor $F$, then the oformula $G$ of $B$ is also won by ${\cal M}_B$, because ${\cal M}_B$ plays in $G$ exactly as ${\cal M}_A$ does. Hence $U_i^{B}$ is won by ${\cal M}_B$. If $G$ is $E$, then its being $\top$-won means that ${\cal M}_B$ wins the $E$ component of $E\vee F$, because ${\cal M}_B$ plays in the $E$ component of $E\vee F$ exactly as ${\cal M}_A$ plays in $E$. Therefore, $E\vee F$ is won by ${\cal M}_B$, and hence so is the $E\vee F$-containing undergroup $U_i^{B}$. The case of $G$ being $F$ is similar.\vspace{2mm}

{\bf (8)}\ Assume $B$ follows from $A$ by Conjunction Introduction. We also assume that the last ($a$'th) oformula of $B$ is $E\wedge F$, and the last two ($a$'th and $b$'th, where $b=a+1$) oformulas of $A$ are $E$ and $F$. As the case of Disjunction Introduction, ${\cal M}_B$ reinterprets every move $a;\vec{w}.\alpha$ (resp. $b;\vec{w}.\alpha$) by either player in the imaginary play of $A$ as the move $a;\vec{w}.1.\alpha$ (resp. $a;\vec{w}.2.\alpha$) by the same player in the real play of $B$, and vice versa.

Let $U_i$ be any undergroup of $B$. If $U_i$ does not contain $E\wedge F$, then the corresponding undergroup $V_i$ of $A$ contains neither $E$ nor $F$. In this case, $U_i$ is won by ${\cal M}_B$ for the same reason as in the preceding clause. If $U_i$ contains $E\wedge F$, then there are two undergroups $V_i^{E}$, $V_i^{F}$ of $A$ corresponding to $U_i$, where $V_i^{E}$ contains $E$ (but not $F$), and $V_i^{F}$ contains $F$ (but not $E$), with all other ($\neq E\wedge F$) oformulas of $U_i$ contained by both $V_i^{E}$ and $V_i^{F}$. Of course, both $V_i^E$ and $V_i^F$ are won by ${\cal M}_A$ because ${\cal M}_A$ wins the overall game $A$. This means that there is an oformula $G_1$ (resp. $G_2$) in $V_i^E$ (resp. $V_i^F$) such that ${\cal M}_A$ wins it. If at least one oformua $G\in\{G_1,G_2\}$ is neither $E$ nor $F$, then the corresponding oformula $G$ of $B$ is won by ${\cal M}_B$, because ${\cal M}_B$ plays in $G$ exactly as ${\cal M}_A$ does. Hence the $G$-containing undergroup $U_i$ of $B$ is won by ${\cal M}_B$. If $G_1$ is $E$ and $G_2$ is $F$, then ${\cal M}_A$ winning them means that ${\cal M}_B$ wins both the $E$ and the $F$ components of $E\wedge F$, because ${\cal M_B}$ plays in the $E$ (resp. $F$) component of $E\wedge F$ exactly as ${\cal M_A}$ does in $E$ (resp. $F$).  Hence $E\wedge F$ is won by ${\cal M}_B$, and hence so is the $E\wedge F$-containing undergroup $U_i$.\vspace{2mm}

{\bf (9)}\ Assume $B$ is obtained from $A$ by Recurrence Introduction. Namely, the last ($a$'th) oformula of $B$ is $\pst F$, and the last ($a$'th) oformula of $A$ is $F$. We further assume that the number of overgroups of $B$ is $n$, and thus the number of overgroups of  $A$ is  $n+1$. In what follows,  $\vec{w}$ is any sequence of $n$ natural numbers, and $b$ is a positive integer not exceeding the number of oformulas of either cirquent.
If $b\neq a$, then ${\cal M}_B$ simply reinterprets every move $b;\vec{w},0.\alpha$ by either player in the imaginary play of $A$ as the move $b;\vec{w}.\alpha$ by the same player in the real play of $B$, and vice versa. If $b=a$, then ${\cal M}_B$ reinterprets, for any positive integer $u$, every move $a;\vec{w},u.\alpha$ by either player in the imaginary play of $A$ as the move $a;\vec{w}.u.\alpha$ by the same player in the real play of $B$, and vice versa.

Consider any undergroup $U_i^{B}$ of $B$. Let $\vec{x}=x_1,\ldots,x_n$ be any sequence of $n$ positive integers. ${\cal M}_A$'s winning $A$ means that $\Gamma_A$ is a $\top$-won run of $A$ and that the corresponding undergroup $U_i^{A}$ of $A$ is won by ${\cal M}_A$. Then, for any sequence $\vec{y}=x_1,\ldots,x_n,x$, where $x$ is any positive integer, there is an oformula $F_b$ in $U_i^{A}$ such that $\Gamma_A^{[b;\vec{y}]}$ is a $\top$-won run of $F_b$. If such $F_b$ is not the $a$'th oformula $F$, then, in the context of $\vec{x}$, the oformula $F_b$ of $B$ is also won by ${\cal M}_B$, i.e. $\Gamma_B^{[b;\vec{x}]}$ is a $\top$-won run of $F_b$, because ${\cal M}_B$ plays in the copy $\vec{x}$ of $F_b$ in $B$ exactly as ${\cal M}_A$ does in the copy $\vec{y}$ of $F_b$ in $A$. Hence $U_i^{B}$ is won by ${\cal M}_B$. If $F_b$ is the $a$'th oformula $F$, then, in the context of $\vec{x}$, the corresponding oformula $\pst F$ of $B$ is won by ${\cal M}_B$ as well, i.e. $\Gamma_B^{[a;\vec{x}]}$ is a $\top$-won run of $\pst F$. This is so because ${\cal M}_B$ plays in the $x$'th component $F$ of the copy $\vec{x}$ of $\pst F$ exactly as ${\cal M}_A$ does in the copy $\vec{y}$ of $F$ in $A$. Namely, $(\Gamma_B^{[a;\vec{x}]})^{x.}=\Gamma_A^{[a;\vec{y}]}$. Since $\Gamma_A^{[a;\vec{y}]}$ is a $\top$-won run of $F$, so is $(\Gamma_B^{[a;\vec{x}]})^{x.}$. Further, due to the arbitrariness of $x$, $\Gamma_B^{[a;\vec{x}]}$ is a $\top$-won run of $\pst F$. Therefore, the $\pst F$-containing undergroup $U_i^{B}$ is won by ${\cal }M_B$.\vspace{2mm}

{\bf (10)}\ Finally, assume that $B$ is obtained from $A$ by Corecurrence Introduction. Let us assume that the last ($a$'th) oformula of $B$ is $\pcost F$, and the last ($a$'th) oformula of $A$ is $F$. And assume that $n$ $(n\geq 0)$ is the number of the {\em new} overgroups $U_j$ in which the $a$'th oformula $F$ was included when moving from $B$ to $A$. Let us further assume that all of such $n$ overgroups are at the end of the list of overgroups of either cirquent. In what follows, let $\vec{w}$ be any sequence of $m$ natural numbers, where $m$ is the total number of overgroups of either cirquent minus $n$. We construct the EPM ${\cal M}_B$ as follows.

Let $f$ be some standard injective function from the set of $n$-tuples $(u_1,\ldots,u_n)$ of positive integers onto the set of positive integers $u$.  In its simulation routine, ${\cal M}_B$ reinterprets every move $a;\vec{w},u_1,\ldots,u_n.\alpha$ made by ${\cal M}_A$ in the imaginary play of $A$ as the move $a;\vec{w},0,\ldots,0.u.\alpha$ ($n$ occurrences of $0$ after $\vec{w}$) in the real play of $B$, where $u=f(u_1,\ldots,u_n)$. Whenever the environment makes a move $a;\vec{w},0,\ldots,0.v.\beta$ (also $n$ occurrences of $0$ after $\vec{w}$) for some positive integer $v$ in the real play of $B$, if there is no $n$-tuple $(u_1,\ldots,u_n)$ such that $v=f(u_1,\ldots,u_n)$, then ${\cal M}_B$ simply ignores it; if $v=f(u_1,\ldots,u_n)$, then ${\cal M}_B$ translates it as the move $a;\vec{w},u_1,\ldots,u_n.\beta$ by ${\cal M}_A$'s adversary in the imaginary play. Note that the above routine works as well in the case of $n=0$. Simply, $f()=c$ for some fixed positive integer $c$, ${\cal M}_B$ reinterprets every move $a;\vec{w}.\alpha$ made by ${\cal M}_A$ in $A$ as the move $a;\vec{w}.c.\alpha$ in $B$, and whenever the environment makes a move $a;\vec{w}.v.\beta$ in $B$, if $v\neq c$, ${\cal M}_B$ ignores it, and if $v=c$, ${\cal M}_B$ translates it as the move $a;\vec{w}.\beta$ by ${\cal M}_A$'s adversary in the imaginary play of $A$.


As usual, consider any undergroup $U_i^{B}$ of $B$, and let $\vec{x}=\vec{y},x_1,\ldots,x_n$ be any sequence of $(m+n)$ positive integers, where $\vec{y}$ is any sequence of $m$ positive integers. Then the corresponding undergroup $U_i^{A}$ of $A$ is won by ${\cal M}_A$, which, in turn, means that there is an oformula $F_b$ ($1\leq b\leq a$) in $U_i^{A}$ such that ${\cal M}_A$ wins it. If such $F_b$ is not the $a$'th oformula $F$, then the corresponding oformula $F_b$ of $B$ is also won by ${\cal M}_B$, because ${\cal M}_B$ plays in $F_b$ of $B$ exactly as ${\cal M}_A$ does in $F_b$ of $A$. Therefore, the $F_b$-containing undergroup $U_i^{B}$ is won by ${\cal M}_B$. If $F_b$ is the $a$'th oformula $F$, then the corresponding oformula $\pcost F$ of $B$ is won by ${\cal M}_B$ as well. This is so because ${\cal M}_B$ plays in at least one component $F$ of $\pcost F$ in $B$ exactly as ${\cal M}_A$ does in $F$ of $A$. Precisely, we have
$(\Gamma_B^{[a;\vec{y},x_1,\ldots,x_n]})^{x.}=\Gamma_A^{[a;\vec{y},x_1,\ldots,x_n]}$, where $x=f(x_1,\ldots,x_n)$. Thus the $\pcost F$-containing undergroup $U_i^{B}$ is won by ${\cal M}_B$.
\end{proof}

\begin{theorem}\label{mainth2}
Every cirquent provable in {\bf CL15}$(\pst)$ is uniformly valid.

Furthermore, there is an effective procedure that takes an arbitrary {\bf CL15}$(\pst)$-proof of an arbitrary cirquent $C$ and constructs a uniform solution of $C$.
\end{theorem}

\begin{proof}
Immediately  from Theorem \ref{mainth1} by induction on the lengths of {\bf CL15}$(\pst)$-proofs.
\end{proof}

\begin{theorem}\label{mainth3}
For any formula $F$, if {\bf CL15}$(\pst)\vdash F$, then $F$ is uniformly valid.

Furthermore, there is an effective procedure which takes any {\bf CL15}$(\pst)$-proof of any formula $F$ and constructs a uniform solution of $F$.
\end{theorem}

\begin{proof}
Immediately from Theorem \ref{mainth2} and Lemma \ref{apr14b}.
\end{proof}

Below, a {\bf uniformly valid $(\neg,\wedge,\vee,\pst,\pcost)$-principle} means the result of replacing every occurrence of the operator $\pst$ (resp. $\pcost$) by the symbol $!$ (resp. $?$) in some uniformly valid $(\neg,\wedge,\vee,\pst,\pcost)$-formula. Similarly, a {\bf uniformly valid $(\neg,\wedge,\vee,\st,\cost)$-principle} means the result of replacing every occurrence of the operator $\st$ (resp. $\cost$) by the symbol $!$ (resp. $?$) in some uniformly valid $(\neg,\wedge,\vee,\st,\cost)$-formula. The reason for introducing these technical concepts is merely to make it possible to directly compare the otherwise syntactically nonidentical $(\neg,\wedge,\vee,\pst,\pcost)$-formulas with $(\neg,\wedge,\vee,\st,\cost)$-formulas.

\begin{theorem}\label{superset}
The set of uniformly valid $(\neg,\wedge,\vee,\pst,\pcost)$-principles is a proper superset of the set of uniformly valid $(\neg,\wedge,\vee,\st,\cost)$-principles.
\end{theorem}

\begin{proof} The fact that the set of uniformly valid $(\neg,\wedge,\vee,\pst,\pcost)$-principles is a {\em superset} of the set of uniformly valid  $(\neg,\wedge,\vee,\st,\cost)$-principles is immediate from Theorems \ref{niu} and \ref{mainth3}. Furthermore, the former set is in fact  a {\em proper} superset of the latter set because, as proven in \cite{Japsep}, the formula $P\wedge\pst(P\rightarrow P\wedge P)\rightarrow\pst P$ is uniformly valid while its counterpart $P\wedge\st(P\rightarrow P\wedge P)\rightarrow\st P$ is not.
\end{proof}

\section{A secondary result}
Japaridze \cite{Japfin,Japfour} claimed that $\st$ is strictly stronger than $\pst$ (and thus $\cost$ is strictly weaker than $\pcost$) in the sense that the formula $\st P\rightarrow\pst P$ is uniformly valid while its converse $\pst P\rightarrow\st P$ is not. The first part of this claim was proven in  \cite{Japtur}, but the second part has never been verified. In order to make our investigation of the relationship between the two sorts of recurrences more comprehensive, below we provide such  a verification.

\begin{theorem}\label{secondary}
The formula $\pst P\rightarrow\st P$ is not uniformly valid.
\end{theorem}

\begin{proof}
Let ${\cal M}$ be an arbitrary EPM, i.e. strategy of the machine $(\top)$. Below we construct a counterstrategy ${\cal C}$ such that, when the environment $(\bot)$ follows it, ${\cal M}$ loses $\pst P\rightarrow\st P$ with $P$ interpreted as a certain enumeration game. Here, an {\bf enumeration game} (\cite{Japsep}) is a game where any natural number, identified with its decimal representation, is a legal move by either player at any time (and there are no other legal moves). It should be noted that, as shown in \cite{Japtam2}, every enumeration game is static, and hence is a legitimate value of an interpretation $^*$ on any atom.  Hence, due to the arbitrariness of ${\cal M}$, $\pst P\rightarrow\st P$ (i.e. $\pcost\neg P\vee\st P$) is not uniformly valid.

Since $P$ is going to be interpreted as an enumeration game and its legal moves are known even before we actually define that interpretation, in certain contexts we may identify formulas with games without creating any confusion. The work of ${\cal C}$ consists in repeating the following interactive routine over and over again (infinitely many times), where $i$ is the number of the iteration. In our description below, a {\em fresh number} means a natural number that  has not yet been chosen in the play by either player as a move in any thread/copy of  $P$.\vspace{2mm}

LOOP($i$): Whenever permission is granted by the machine ${\cal M}$, make the move $2.w.u$, where $u$ is a fresh number and $w$ is the $i$th finite bitstring of the lexicographic list of all finite bitstrings.
\vspace{2mm}


Consider the run $\Delta$ generated by $\cal M$ in the scenario when its adversary follows the above counterstrategy. Let $\Omega=\Delta^{1.}$ and $\Gamma=\Delta^{2.}$. That is, $\Omega$ is the (sub)run that took place in the $\pcost\neg P$ component, and $\Gamma$ is the (sub)run that took place in the $\st P$ component. From some analysis of the work of LOOP, details of which are left to the reader, one can see that  $\Gamma^{\preceq x_1}\neq\Gamma^{\preceq x_2}$ for any two different infinite bitstrings $x_1$ and $x_2$. Hence, as there are uncountably many infinite bitstrings while only countably many positive integers, there is an infinite bitstring $y$ such that, for every positive integer $v$,  $\Omega^{v.}\neq\neg\Gamma^{\preceq y}$. Fix this $y$.

Now we select an interpretation $^*$ that interprets $P$ as the enumeration game such that, for any legal run $\Theta$ of the game $P$, $Wn^{P}\langle\Theta\rangle=\bot$ iff $\Theta=\Gamma^{\preceq y}$. We claim that ${\cal M}$ loses the overall game under this interpretation. First, it is obvious that ${\cal M}$ loses the game $P$ in the thread $y$, which means that it loses the $\st P$ component. Next, ${\cal M}$ also loses the $\pcost\neg P$ component because it loses every component $\neg P$ of $\pcost\neg P$. This is so because the run that took place in any component $\neg P$ of $\pcost\neg P$ is won by $\top$ iff it is $\neg\Gamma^{\preceq y}$, which, however, is impossible (due to the above analysis).
\end{proof}

An alternative albeit non-constructive and less direct proof of Theorem \ref{secondary} would rely on Theorem \ref{superset}. Namely, one could show that, if  $\pst P\rightarrow\st P$ was uniformly valid and hence (in view of the already known fact of the uniform validity of the converse of this formula) $\st P$ and $\pst P$ were ``logically equivalent'', then they would induce identical logics, in the precise sense that the set of uniformly valid $(\neg,\wedge,\vee,\pst,\pcost)$-principles would coincide with the set of uniformly valid $(\neg,\wedge,\vee,\st,\cost)$-principles, contrary to what Theroem \ref{superset} asserts.

\end{document}